\documentclass[a4paper]{jpconf}

\usepackage{graphicx}
\usepackage{subfigure}

\begin{document}

%%%%%%%%%%%%%%%%%%%%%%%%%%%%%%%%%%%%%%%%%%%%%%%%%%%%%%%%%%%%%%%%%%%%%%%%%%%%%%%%%%%%%%%%%%%%%%%%
\title{Scintillation balloon for liquid scintillator base 
 Neutrinoless double beta decay search experiments} %\jpcs}

%%%%%%%%%%%%%%%%%%%%%%%%%%%%%%%%%%%%%%%%%%%%%%%%%%%%%%%%%%%%%%%%%%%%%%%%%%%%%%%%%%%%%%%%%%%%%%%%
\author{S~Obara\footnote{Current address; Department of Physics, Kyoto University, Kyoto 606-8502, Japan}, Y~Gando, and K~Ishidoshihro}
\address{Research Center for Neutrino Science, Tohoku University, Sendai 980-8578, Japan}
\ead{obara@awa.tohoku.ac.jp}

%%%%%%%%%%%%%%%%%%%%%%%%%%%%%%%%%%%%%%%%%%%%%%%%%%%%%%%%%%%%%%%%%%%%%%%%%%%%%%%%%%%%%%%%%%%%%%%%
\begin{abstract}
A liquid scintillator base experiment KamLAND-Zen has set a lower limit on neutrinoless double beta decay half-life, and upgrade project KamLAND-Zen 800 has started in 2019. 
Unfortunately this project expects some backgrounds, and one of the main backgrounds is $\beta$/$\gamma$-ray from ${}^{214}$Bi in container of xenon loaded liquid scintillator (mini-balloon). 
In order to reject the background, we suggest using scintillation film for the future mini-balloon. 
If we can tag $\alpha$-ray from ${}^{214}$Po by scintillation detection, we can eliminate ${}^{214}$Bi events by delayed coincidence analysis.

Recently, it was reported that polyethylene naphthalate (PEN) can be used as a scintillator with blue photon emission. 
PEN has chemical compatibility for strong solvent, thus it has a possibility to use in liquid scintillator. 
In this presentation, we will mention the results for feasibility studies about transparency and emission spectra, light yield, radioactivity, strength of film etc.. 
We also show the test-sized scintillation balloon with an 800-mm diameter and discussions about how to use the scintillation balloon in KamLAND.

\end{abstract}

%%%%%%%%%%%%%%%%%%%%%%%%%%%%%%%%%%%%%%%%%%%%%%%%%%%%%%%%%%%%%%%%%%%%%%%%%%%%%%%%%%%%%%%%%%%%%%%%
\section{Introduction}
Neutrinoless double-beta decay ($0\nu2\beta$) has an important role to solve the Majorana nature of neutrinos, but it is still not observed yet.
KamLAND-Zen 400 and GERDA searched and reported the lower limit of its half-life as about $1 \times 10^{26}$~yr \cite{klz400, gerda}.

In the KamLAND-Zen 800 experiment as an upgrade program of KamLAND-Zen 400, newly fabricated mini-balloon is installed that is made of extremely low-background nylon film and holds 745 kg of 91\% enriched ${}^{136}$Xe loading an ultra-pure liquid scintillator. 
However, the backgrounds still remain; the dominant background is $\beta$/$\gamma$-rays ($Q_{\beta/\gamma}$ = 3.2~MeV) from ${}^{214}$Bi contamination in the nylon film.

\begin{figure}[htbp]
	\centering
	\includegraphics[width=0.5\linewidth]{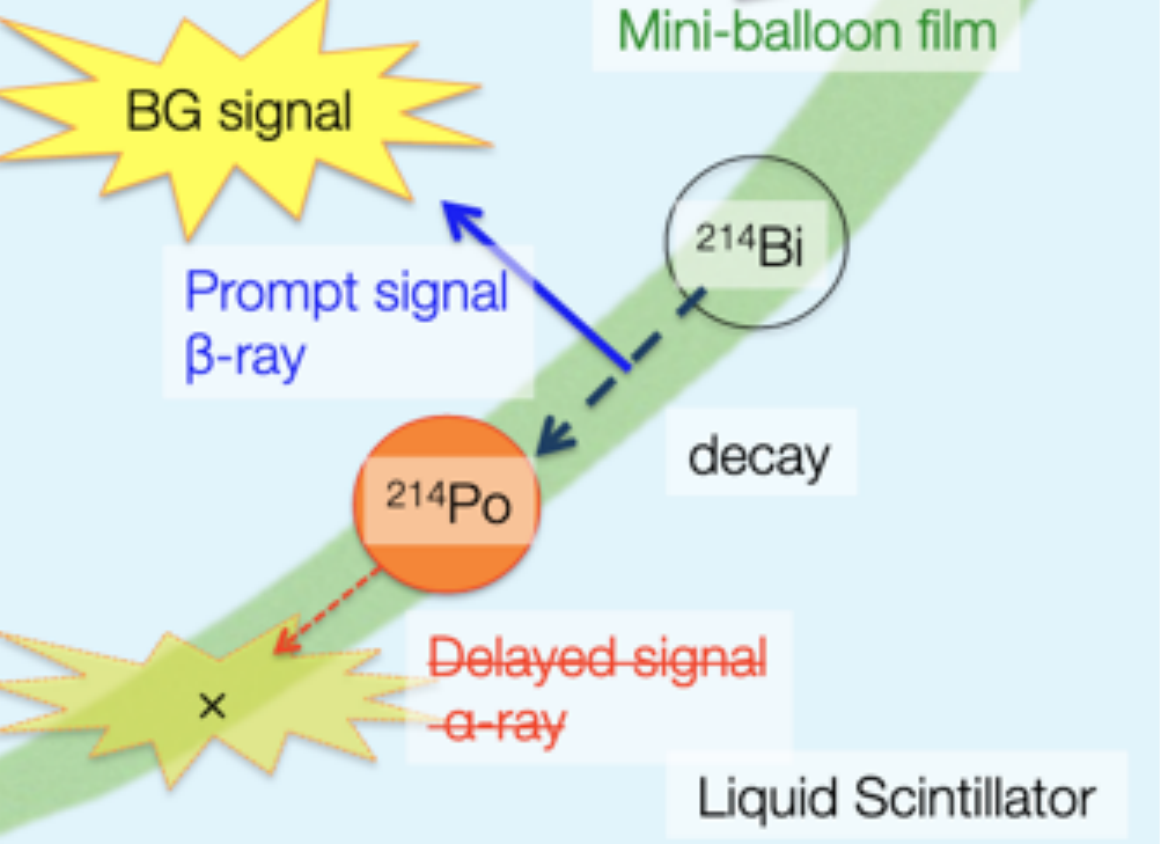}
	\caption{Schematic view of a sequential decays of ${}^{214}$Bi--${}^{214}$Po in the inner-balloon film material. $\beta$/$\gamma$-rays from ${}^{214}$Bi easily contributes as a background in a liquid-scintillator region but $\alpha$-rays from ${}^{214}$Po sometimes stop in a nylon film. In such case, ${}^{214}$Bi-$\beta$/$\gamma$ remains as a background without delayed-coincidence veto. (color online)}
	\label{Fig:BiPo}
\end{figure}

In order to identify ${}^{214}$Bi background, delayed-coincidence method is useful with a tagging its sequential decays of ${}^{214}$Bi--${}^{214}$Po ($Q_{\alpha}$=7.8~MeV, $T_{1/2}$=164.3~$\mu$sec) and its tagging efficiency is 99.97\% in the liquid-scintillator.
Unfortunately, at the near of the mini-balloon, about 50\% of ${}^{214}$Po $\alpha$-ray stops in the nylon film material because of its short mean free path, and the tagging efficiency also decreases (Figure~\ref{Fig:BiPo}).

Here we propose using a PolyEthyleneNaphthalate (PEN) film as the mini-balloon material in order to detect all of the ${}^{214}$Po $\alpha$-ray for ${}^{214}$Bi background rejection with the delayed-coincidence.
PEN is recently reported it has a blue photon emission \cite{pen} and it has commercially high chemical resistance, flexibility, and high transparency.
We have developed feasibility studies of a PEN-based mini-balloon (scintillation balloon) for a future KamLAND-Zen and found that it can identify 99.7\% of ${}^{214}$Bi backgrounds \cite{scintiballoon}. 
Currently, we started to fabricate a test-size scintillation balloon and develop a detail of its fabrication and handling for introducing into a future KamLAND2-Zen project.

This scintillation balloon is an essential vessel for a liquid-scintillator detector to suppress the radioactive backgrounds.
In addition, recently some experiments also plan to use a PEN as self-vetoing structural material~\cite{penvessel} or as a liquid-argon vessel~\cite{penar}.
PEN has the various potential for using in low-background required experiments.

%%%%%%%%%%%%%%%%%%%%%%%%%%%%%%%%%%%%%%%%%%%%%%%%%%%%%%%%%%%%%%%%%%%%%%%%%%%%%%%%%%%%%%%%%%%%%%%%
\section{Test Fabrication and Heat-Welding with PEN}
Both of current nylon-based mini-balloon in the KamLAND-Zen 400/800 are fabricated by heat-welding with 24-gores and some parts.
Welding parameters were optimized for a 25-$\mu$m thickness nylon film, but it is not cleared for a PEN film.

\begin{figure}[htbp]
    \subfigure[Schematic view of measurement]{
		\includegraphics[width=0.3\columnwidth]{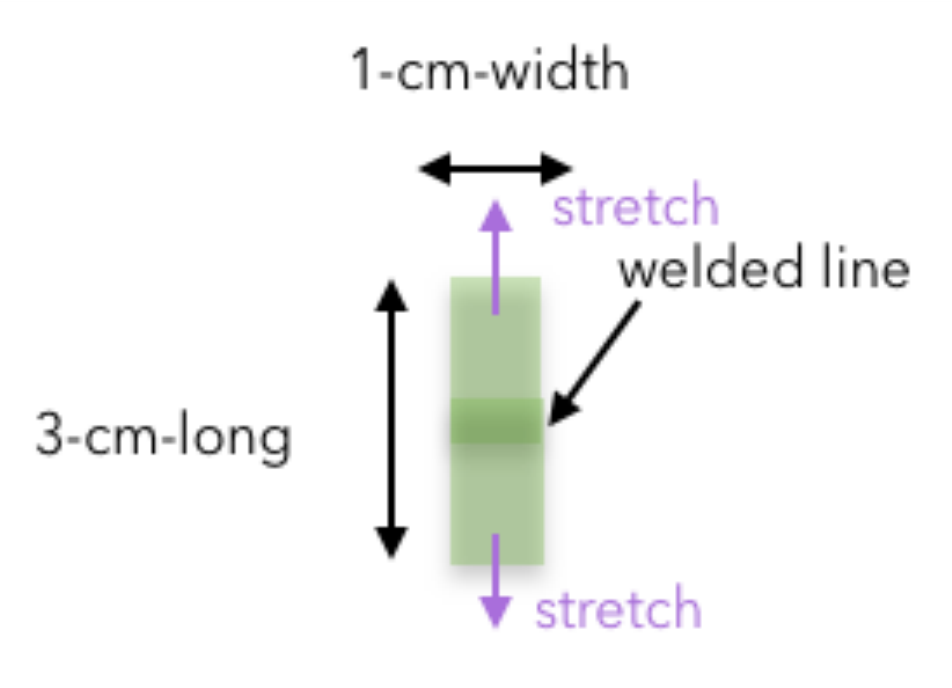}
	}
	\subfigure[Result of film strength]{
		\includegraphics[width=0.4\columnwidth]{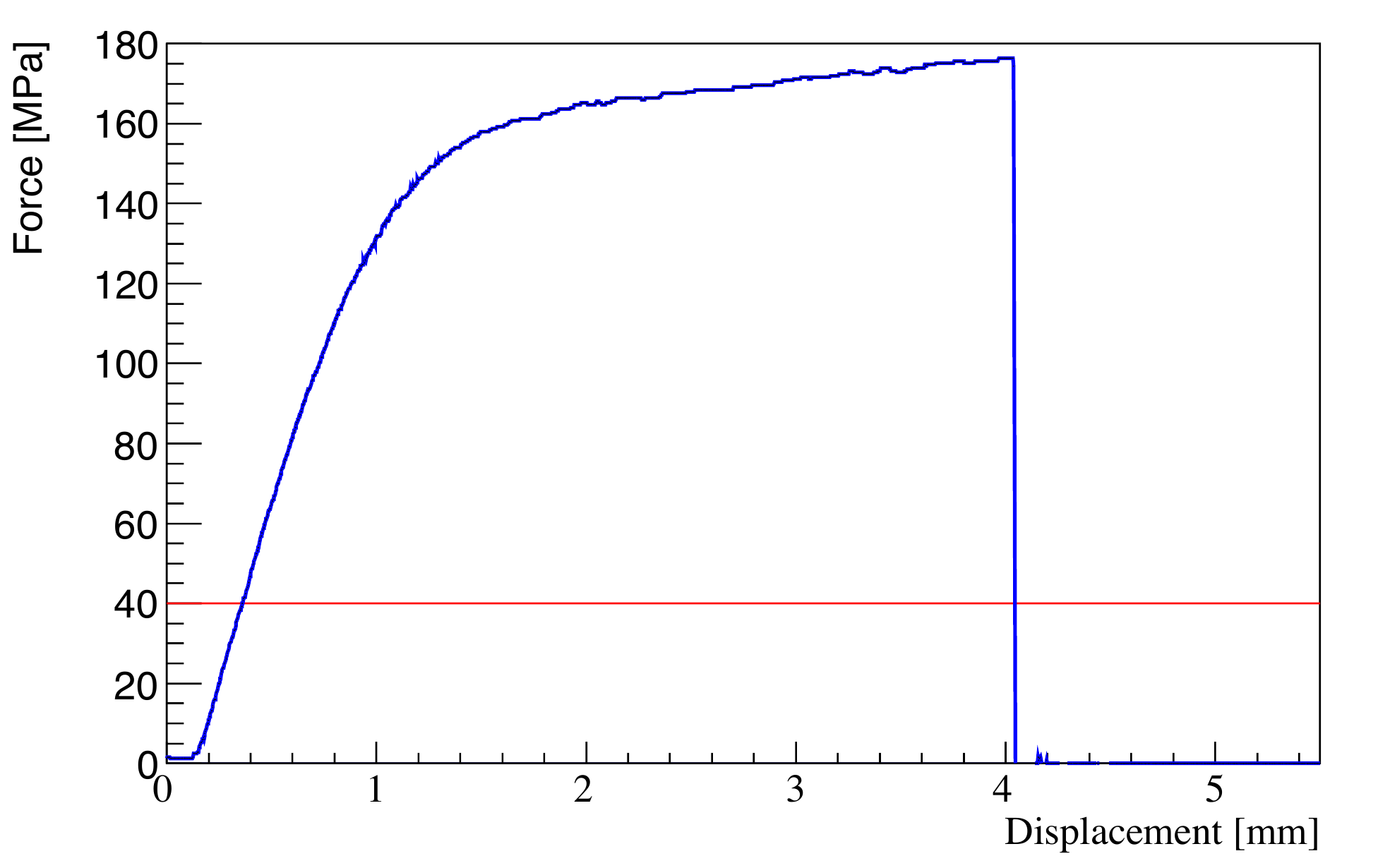}	
	}
	\subfigure[A test-size balloon]{
		\includegraphics[bb=0 0 496 752, width=0.2\linewidth]{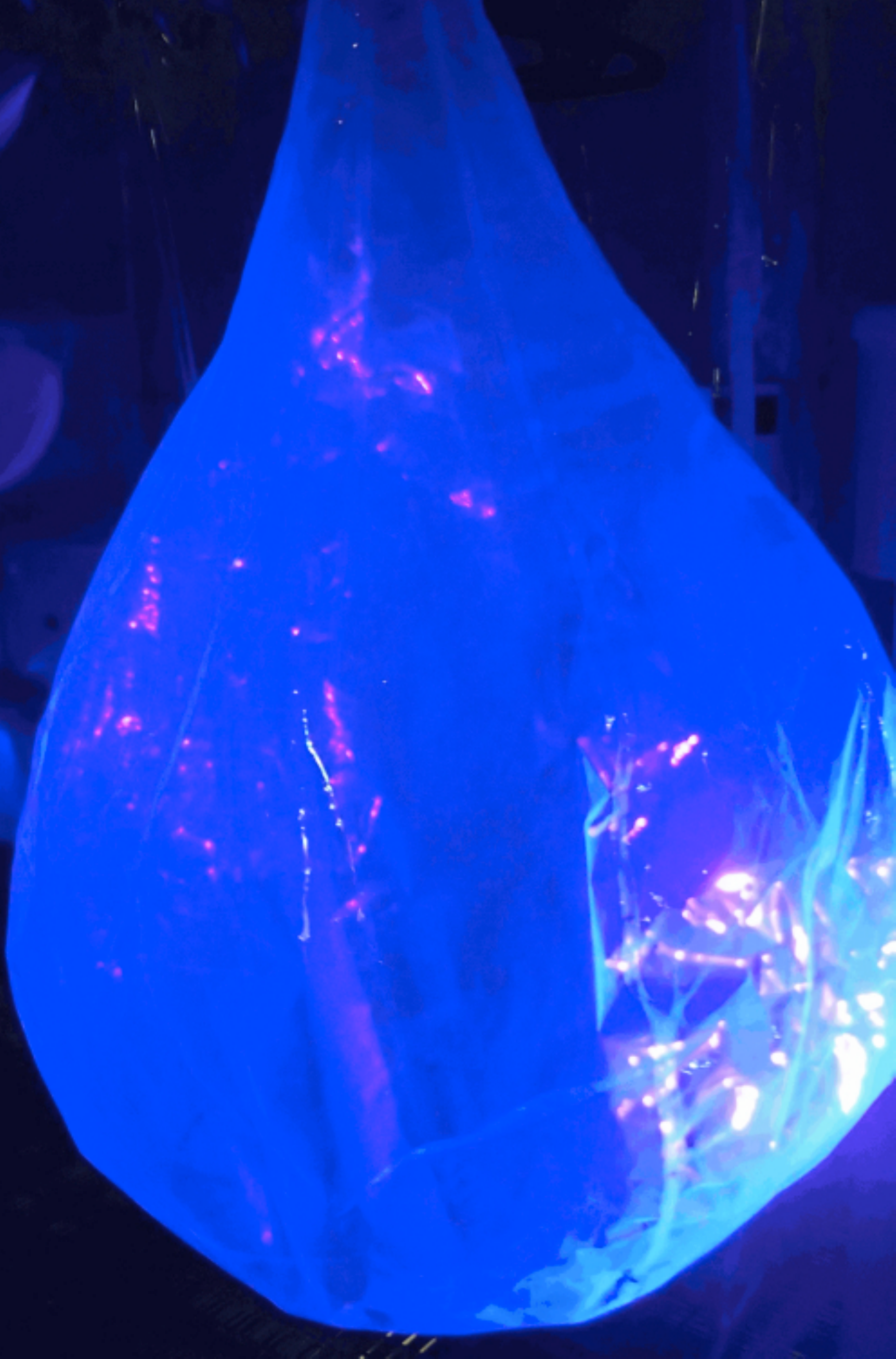}
	}
	\caption{Film strength measurement for the heat-welding line using a force gauge IMADA ZTA-500. The horizontal axis corresponds to the stretching of the film. The sample size is a 1-cm width and 3-cm long for 25-$\mu$m thickness, with a welding line at the center of it. The red horizontal line is a lower requirement strength of 40~MPa. A test-size scintillation balloon with ultraviolet light shows blue photon emission. (color online)}
	\label{Fig:OnWeldingLine}
\end{figure}

We searched about welding parameters with satisfying with our requirement of 40~MPa strength on a welded line in order to hold the liquid scintillator.
The welded line is approximately 0.5~cm overlapped.
The parameters are tuned as follows; maximum temperature 220~${}^\circ\mathrm{C}$, heating time 2.0~s, and cooling temperature 80~${}^\circ\mathrm{C}$, but these parameters may depend on the humidity so that we have to re-tune the parameters {\it in-situ} environment.
Film strength measurement of its welded PEN film by a force gauge (IMADA ZTA-500) is shown in Figure~\ref{Fig:OnWeldingLine}. 
A breaking point is about 175~MPa strength with 4.05~mm displacement and it is satisfied our requirement.
We also fabricated an 800-mm diameter test-size of PEN-based scintillation balloon for checking its handing compared with nylon film.

Since the film is very thin 25~$\mu$m, the fabricated balloon has sometimes tiny holes made by the heat-welding.
Therefore we have to check it with helium-gas leakage after fabrication.
In a nylon-balloon case, such tiny holes can be repaired by a patch-film with adhesive.
However in the PEN film case, almost all adhesive cannot be used because of high chemical resistance of PEN.
The current issue is such tiny hole repairing for PEN-based balloon although re-heating may be applicable.

%%%%%%%%%%%%%%%%%%%%%%%%%%%%%%%%%%%%%%%%%%%%%%%%%%%%%%%%%%%%%%%%%%%%%%%%%%%%%%%%%%%%%%%%%%%%%%%%
\section{Summary}
We proposed a PEN-based scintillation balloon for a liquid-scintillator type detector, particularly a future KamLAND2-Zen project, in order to distinguish ${}^{214}$Bi background for $0\nu2\beta$ search.
Its test-size fabrication was started and we conduct the optimization of heat-welding parameters for a PEN film.

%%%%%%%%%%%%%%%%%%%%%%%%%%%%%%%%%%%%%%%%%%%%%%%%%%%%%%%%%%%%%%%%%%%%%%%%%%%%%%%%%%%%%%%%%%%%%%%%

\end{document}